\documentclass[aps,prb,reprint,superscriptaddress]{revtex4-2}

\usepackage{amsmath,amssymb,bm,graphicx}
\usepackage[T1]{fontenc}
\usepackage[utf8]{inputenc}

\usepackage[
    colorlinks=true,
    linkcolor=blue,
    citecolor=blue,
    urlcolor=blue
]{hyperref}
\usepackage{orcidlink}

\begin{document}

\title{Continuum  Dirac Theory and Nonlinear Extensions of a Chain of Alternating non-linear springs: an SSH version of the Fermi-Pasta-Ulam-Tsingou model}

\author{Gerardo G. Naumis\,\orcidlink{0000-0002-1338-1522}}
\email{naumis@fisica.unam.mx}

\author{Ilder I. Salgado\,\orcidlink{0009-0009-1778-4810}}
\email{ilder@estudiantes.fisica.unam.mx}

\author{Jorge Fujioka}
\email{fujioka@fisica.unam.mx}
\affiliation{Depto. de Sistemas Complejos, Instituto de Física,
Universidad Nacional Autónoma de México (UNAM),
Av. Universidad 3004, Ciudad Universitaria, Coyoacán,
04510, CDMX, Mexico.}

\date{\today}
\begin{abstract}
A mechanical version of the SSH model, made of equal masses joined by two alternating non-linear springs is studied  in the continuum limit. Such a non-linear dimerized mechanical chain generates a nonlinear, dispersive, two-field hyperbolic system of Boussinesq type of differential equations, describing coupled acoustic and gapped optical fields. The corresponding Lagrangian density is found, revealing  that the acoustic and optical degrees of freedom are coupled through a structure that is formally analogous to a synthetic spin–orbit coupling, where the spatial motion is locked to the internal sublattice degree of freedom. An emergent   
non-Abelian Berry connection term also appears. Some particular solitary-wave solutions are found using variational theory, as the effective optical field obeys an undamped and unforced cubic--quintic Duffing equation. Near the Brillouin-zone edge, the system becomes a nonlinear Dirac field theory of the Soler type, providing a direct mechanical realization of interacting relativistic solitons. By factoring out fast spatial oscillations, an effective Jackiw--Rebbi Hamiltonian is obtained, with topological zero modes at domain walls. These modes are akin to the electronic flat bands in moiré graphene systems associated with superconductivity and charge fractionalization. Finally, we discuss the mechanical rigidity transition and its effects on the fields as the system evolves toward the fully dimerized limit, in which the zero modes are floppy modes.
\end{abstract}

\maketitle

\section{Introduction}

The Su--Schrieffer--Heeger (SSH) model was introduced to describe solitons in polyacetylene, a chain with two alternating transfer integrals produced by a Peierls dimerization~\cite{su1979solitons,heeger1988solitons}. Despite its simplicity, it is the prototypical one-dimensional topological Hamiltonian: its two bands carry a quantized Zak phase (winding number) that changes when the intracell and intercell couplings are exchanged, and it fixes the bulk-boundary correspondence of the chain~\cite{zak1989berry,berg2010bulk,asboth2016short}. 

A mechanical version of the SSH model is obtained from a chain of equal masses joined by alternating springs. Its dynamical matrix has the SSH structure, so the vibrational spectrum consists of a gapless acoustic branch and a gapped optical branch, whose separation is controlled by the contrast between the two stiffnesses. This analogy has been exploited in topological mechanics~\cite{kane2014topological,huber2016topological,susstrunk2016classification}, in acoustic waveguides~\cite{coutant2021acoustic}, and in electrical circuits~\cite{sahin2025topolectrical}. When the contrast is pushed to its limit, the weak springs disappear and the chain breaks into isolated dimers. In the language of rigidity theory, the acoustic modes then become floppy modes, since half of the mechanical constraints are lost~\cite{phillips1979topology,calladine1978,Naumis2005,flores2010heating,mao2018maxwell}. This rigidity transition has a precise electronic counterpart: the flat bands of twisted bilayer graphene at magic angles correspond to floppy modes after a supersymmetry transformation~\cite{flores2010heating, Naumis2021}.

Nonlinearity enters in a natural way in the mechanical chain, because the springs are anharmonic. In this case, the SSH structure is combined with the physics of the Fermi--Pasta--Ulam--Tsingou (FPUT) problem~\cite{fermi1955studies}. The continuum limit of the FPUT chain led to the Korteweg--de Vries equation and to the discovery of solitons~\cite{zabusky1965interaction}, which is a central result of the field (see Refs.~\cite{ford1992fermi,berman2005fermi,campbell2005introduction,gallavotti2007fermi,verhulst2020variations} for reviews). For a monoatomic chain, a gradient expansion of the lattice equations produces a single nonlinear dispersive wave equation of Boussinesq type. Nonlinear effects in SSH-like mechanical lattices have been studied mainly through discrete models, focusing on amplitude-dependent topological edge states and their delocalization~\cite{pal2018amplitude,many2022nonlinear,sone2025,ake2026}. A continuum field theory for the bulk of the nonlinear mechanical SSH chain, which connects it with the Dirac theory above, has not been developed. 

In this work we derive systematically the continuum limit of a nonlinear mechanical SSH chain with equal masses and alternating springs, with quartic bond potentials. We show that it gives rise to a nonlinear, dispersive, two-field system of Boussinesq type for the acoustic and optical fields \cite{ChristovMauginPorubov2007, PorubovAndrianov2013}, with a Lagrangian density in which the gradient coupling between both fields is formally analogous to a synthetic spin--orbit coupling \cite{Zheng2024, Wang2025SpinOrbit, Gu2019SpinOrbit} and to an emergent non-Abelian connection. In the long-wavelength limit, the chain provides a mechanical realization of interacting relativistic solitons. The quartic bond nonlinearity studied here is complementary to the cubic ($\alpha$-type) nonlinearity analyzed in a companion work, where the same chain is used to study how the acoustic--optical gap controls thermalization~\cite{ake2026}.

The layout of this work is the following. In Sec.~II we introduce the dimerized mechanical chain, its linear spectrum and the nonlinear bond forces. In Sec.~III we derive the continuum limit: we obtain the linear acoustic and optical fields, the nonlinear two-field system, the classification of its nonlinear terms and the Lagrangian density. Sec.~IV discusses the long-wavelength limit and the effective acoustic equation. Sec.~V presents the Dirac limit near the zone edge and the zero modes, together with the rigidity transition. The nonlinear Dirac theory is derived in Sec.~VI, and its Soler limit in Sec.~VII. Conclusions are given in Sec.~VIII, and the solitary-wave solutions of the continuum model are collected in Appendix~A.

\section{Dimerized mechanical chain}

\begin{figure}[h!]
    \centering
    \includegraphics[width=\linewidth]{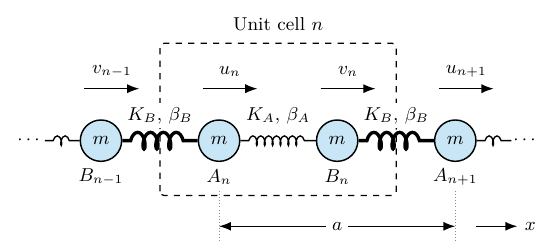}
    \caption{Schematic of the dimerized mechanical chain with equal masses $m$. The dashed box marks unit cell $n$, containing sites $A_n$ and $B_n$, whose longitudinal displacements are $u_n$ and $v_n$, respectively. Intracell and intercell springs are characterized by the linear stiffnesses $K_A$ and $K_B$ and the cubic nonlinear coefficients $\beta_A$ and $\beta_B$. The lattice period $a$ is the distance between equivalent sites in neighboring cells. Different spring drawings distinguish the two bond types.}
\label{fig:dimerized-chain}
\end{figure}

As shown in Fig. \ref{fig:dimerized-chain}, we consider a one-dimensional chain of $N=2M$ identical masses $m$ with periodic boundary conditions. The lattice is divided into $M$ unit cells, each containing two sublattice sites. In equilibrium, the distance between sites is $a/2$. The displacements from the equilibrium position are denoted by,
\begin{equation}
u_n = x_{2n-1}, \qquad v_n = x_{2n}, \qquad n=1,\dots,M.
\end{equation}

Let us now discuss below the properties of this mechanical model. First we start the discussion with linear springs and thereafter we include the effects of non-linearity.

\subsection{Linear Hamiltonian}
 
In the case of the pure linear system, the alternate springs are described by:
\begin{itemize}
\item intracell coupling $K_A$ between $u_n$ and $v_n$,
\item intercell coupling $K_B$ between $v_n$ and $u_{n+1}$.
\end{itemize}

The potential and kinetic energies are
\begin{align}
V &= \frac{1}{2}\sum_{n=1}^{M}\left[
K_A(v_n-u_n)^2 + K_B(u_{n+1}-v_n)^2
\right], \\
T &= \frac{m}{2}\sum_{n=1}^{M}\left(\dot u_n^2 + \dot v_n^2\right).
\end{align}
 
The corresponding Euler--Lagrange equations yield,
\begin{align}
m\ddot u_n &= K_A(v_n-u_n) + K_B(v_{n-1}-u_n), \\
m\ddot v_n &= K_A(u_n-v_n) + K_B(u_{n+1}-v_n).
\end{align}
 
Such equations are easily solved through the use of a Bloch ansatz,
\begin{equation}
u_n = \bar{u}_q e^{i(qna-\omega t)}, \qquad v_n = \bar{v}_q e^{i(qna-\omega t)},
\end{equation}
which leads to the eigenvalue problem
\begin{equation}
D(q)\begin{pmatrix}\bar{u}_q\\\bar{v}_q\end{pmatrix}=\omega^2
\begin{pmatrix}\bar{u}_q\\\bar{v}_q\end{pmatrix},
\end{equation}
with the dynamical matrix defined as,
\begin{equation}
D(q)=\frac{1}{m}
\begin{pmatrix}
K_A+K_B & -(K_A+K_B e^{-iqa}) \\
-(K_A+K_B e^{iqa}) & K_A+K_B
\end{pmatrix}.
\end{equation}
 The spectrum is given by the eigenvalues of $D(q)$,
\begin{equation}
\omega_\pm^2(q)=\frac{K_0}{m}
\pm \frac{1}{m}\sqrt{K_0^2-4K_AK_B\sin^{2}\left(\frac{qa}{2}\right)}
\label{eq:lineardispersion}
\end{equation}
where we defined an effective elastic constant,
\begin{equation}
    K_0=K_A+K_B.
\end{equation}
The $-$ branch corresponds to the acoustic mode, whereas the $+$ branch corresponds to the optical mode,. This is formally identical to the SSH model, developed to understand the Peierls instability of CH$_x$ and later on considerered as the most simple topological Hamiltonian~\cite{su1979solitons, heeger1988solitons, asboth2016short}. In the present case, the topology is  encoded in the function,
\begin{equation}
f(q)=K_A+K_B e^{-iqa}.
\end{equation}

\begin{figure}[h]
    \centering
    \includegraphics[width=1\linewidth]{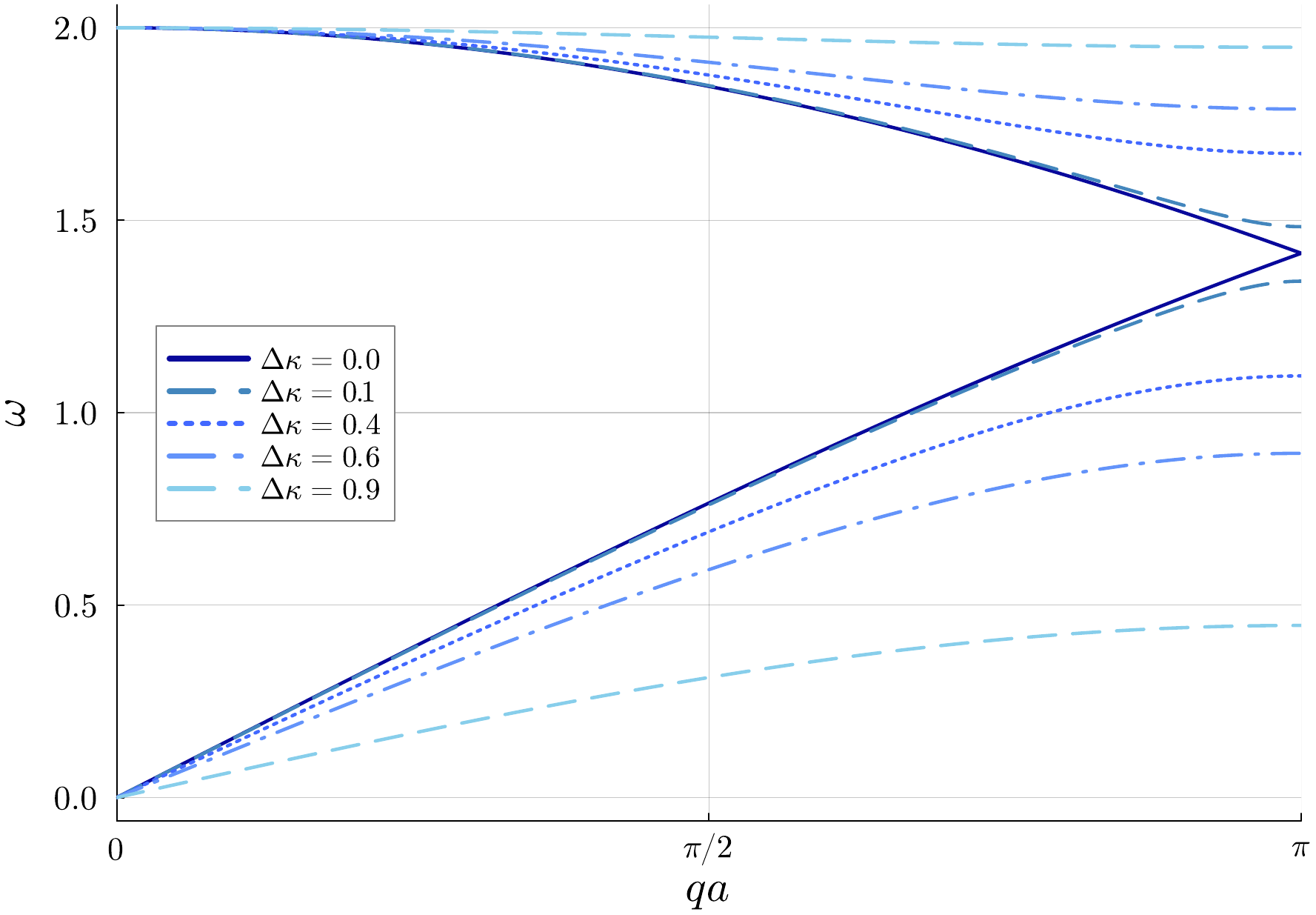}
    \caption{Dispersion relation of the discrete dimerized chain [Eq. \ref{eq:lineardispersion}], with
$K_A=\bar K(1+\Delta\kappa)$ and
$K_B=\bar K(1-\Delta\kappa)$.
The lower and upper branches correspond to the acoustic and
optical modes, respectively. Increasing the coupling contrast
widens the frequency gap at the Brillouin-zone boundary
$qa=\pi$ and reduces the acoustic-branch slope near $q=0$.
For equal couplings ($\Delta\kappa=0$), the two branches meet
at the zone boundary.}
    \label{fig:placeholder}
\end{figure}
 
\subsection{Nonlinear mechanical SSH chain}
 
We extend the model by including quartic bond potentials,
\begin{equation}
V_{ij}=\frac{K_{ij}}{2}(x_j-x_i)^2+\frac{\beta_{ij}}{4}(x_j-x_i)^4,
\label{eq:nonlinearV}
\end{equation}
with alternating coefficients $(K_A,K_B)$ and $(\beta_A,\beta_B)$. The equations of motion become
\begin{align}
m\ddot u_n &= F_A(v_n-u_n)+F_B(v_{n-1}-u_n), \\
m\ddot v_n &= F_A(u_n-v_n)+F_B(u_{n+1}-v_n),
\label{eq:MotionNonLinear}
\end{align}
where $F_{A/B}(\Delta)$ is the function,
\begin{equation}
F_{A/B}(\Delta)=K_{A/B}\Delta+\beta_{A/B}\Delta^3.
\label{eq:ForceNonLinear}
\end{equation}
 
\section{Continuum limit}
 
We introduce the continuum coordinate $x=na$,  and smooth fields
\begin{equation}
u_n(t)\to u(x,t), \qquad v_n(t)\to v(x,t).
\end{equation}
Then we perform a gradient expansion to second order,
\begin{align}
u_{n+1} &\approx u + a u_x + \frac{a^2}{2}u_{xx}, \\
v_{n-1} &\approx v - a v_x + \frac{a^2}{2}v_{xx}.
\end{align}
 
\subsection{Linear continuum theory}
 
For the linear Hamiltonian we obtain the following continuum equations,
\begin{align}
m u_{tt} &= K_0(v-u) - aK_B v_x + \frac{a^2K_B}{2}v_{xx}, \\
m v_{tt} &= K_0(u-v) + aK_B u_x + \frac{a^2K_B}{2}u_{xx}.
\end{align}
 
Now we introduce  acoustic and optical fields,
\begin{equation}
U=\frac{u+v}{2}, \qquad W=\frac{v-u}{2}.
\end{equation}
The continuum fields $U(x,t)$ and $W(x,t)$ have clear physical interpretations. The field
$U(x,t)$ describes the acoustic (center-of-mass) mode, while $W(x,t)$ corresponds to the
optical (relative displacement) mode associated with the dimer deformation, i.e., the intracell stretch.
Taking half the sum and half the difference of the two equations above (with $u=U-W$, $v=U+W$),
the linear continuum equations become,
\begin{align}
    mU_{tt} &= -aK_B W_x + \frac{a^2K_B}{2}U_{xx}
    , \label{Eq:continuum_linear_1}\\
    mW_{tt} &= -2K_0 W + aK_B U_x
    - \frac{a^2K_B}{2}W_{xx}. \label{Eq:continuum_linear_2}
\end{align}
In this limit, one obtains a dispersive wave equation for the acoustic field $U(x,t)$, coupled to a kind of Helmholtz time dependent equation, elliptical differential equation, for the optical field $W(x,t)$.  The system constitutes a two-field continuum theory derived from a dimerized lattice, where the fields play distinct dynamical roles. The field $U(x,t)$ behaves as a
sound-wave (acoustic) excitation, whereas $W(x,t)$ describes a gapped internal oscillator
associated with relative motion within each unit cell. A key feature of the dynamics is the presence of the term $-2K_0 W$, which acts as a mass
term for the optical field. As a consequence, $W$ is a massive field, while $U$ remains
effectively massless and governs the low-energy acoustic sector of the theory.
The coupling terms $-aK_BW_x$ and $+aK_BU_x$ have the same coefficient and opposite signs
(an antisymmetric coupling), and the negative sign of the $W_{xx}$ term
reflects the downward curvature of the optical branch of Eq.~(9) near $q=0$.

For the linear case, we seek solutions of the form,
\begin{equation}
U(x,t)=U_q e^{i(qx-\omega t)},
\qquad
W(x,t)=W_q e^{i(qx-\omega t)}.
\end{equation}
with $U_q$ and $W_q$ two constants determined from the initial and boundary conditions.
Substitution in the corresponding differential equations we obtain, the following
\begin{equation}
\begin{pmatrix}
-m\omega^2+\dfrac{a^2K_Bq^2}{2}
&
i aK_Bq
\\[6pt]
-i aK_Bq
&
-m\omega^2+2K_0-\dfrac{a^2K_Bq^2}{2}
\end{pmatrix}
\begin{pmatrix}
U_q\\
W_q
\end{pmatrix}
=0.
\end{equation}
 
For a nontrivial solution, the determinant must vanish,
\begin{equation}
\left(m\omega^2-\frac{a^2K_Bq^2}{2}\right)
\left(m\omega^2-2K_0+\frac{a^2K_Bq^2}{2}\right)=a^2K_B^2q^2 ,
\end{equation}
leading to,
\begin{equation}
\omega_\pm^2(q)
=
\frac{1}{m}
\left[
K_0
\pm
\sqrt{
K_0^2-a^2K_AK_B\,q^2+\frac{a^4K_B^2}{4}q^4
}
\right].
\qquad
\end{equation}
 
The previous results coincides with Eq. (\ref{eq:lineardispersion}) except for the coefficient in front of the $q^{4}$ term, a difference that comes out from the continuum approximation truncation. The general solution of the linear case can be written as
\begin{align}
U(x,t)
&=
\sum_{s=\pm}
\int dq\,
A_s(q)
e^{i[qx-\omega_s(q)t]}
+\mathrm{c.c.},
\\[4pt]
W(x,t)
&=
\sum_{s=\pm}
\int dq\,
\bigl(-i\bigr)R_s(q)A_s(q)
e^{i[qx-\omega_s(q)t]}
+\mathrm{c.c.},
\end{align}
where
\begin{equation}
R_s(q)
=
\frac{
m\omega_s^2(q)
-\dfrac{a^2K_Bq^2}{2}
}{
aK_Bq
}.
\end{equation}
 
The functions $A_\pm(q)$ are determined by the initial conditions.
For $q\rightarrow0$, the acoustic branch has the expansion
\begin{equation}
\omega_-^2(q)
\simeq
\frac{a^2K_AK_B}
{2m(K_A+K_B)}
q^2,
\end{equation}
which coincides with the long-wavelength limit of the exact lattice branch. Consequently,
\begin{equation}
\omega_-(q)\simeq c|q|
\end{equation}
with sound velocity
\begin{equation}
c^2=
\frac{a^2K_AK_B}
{2m(K_A+K_B)}
.
\end{equation}
 
In this limit $m\omega^2\to mc^2q^2$ and $R_-(q)\simeq -\,aK_Bq/(2K_0)$, so that
\begin{equation}
W_q
\simeq
i\frac{aK_B}{2K_0}\,qU_q .
\end{equation}
The same result follows directly from Eq.~(\ref{Eq:continuum_linear_2}) by neglecting
$W_{tt}$ and $W_{xx}$, which gives $2K_0W\simeq aK_BU_x$.
Transforming back to real space gives, to leading order in spatial
gradients,
\begin{equation}
W
\simeq
\frac{aK_B}{2K_0}U_x
=
\frac{aK_B}{2(K_A+K_B)}U_x .
\end{equation}
This is the static force balance of the two bonds in series: the intracell stretch $v-u=2W$
carries the fraction $K_B/(K_A+K_B)$ of the total elongation $aU_x$ of the cell.
 
Thus, in the long-wavelength limit, the optical field $W$ is slaved to
the spatial strain of the acoustic field $U$. The optical field is
therefore not an independent low-energy degree of freedom, and it can
be eliminated perturbatively to obtain an effective equation for the
acoustic field.

\subsection{Nonlinear contribution}
 
Using Eq. (\ref{eq:ForceNonLinear}), we need the arguments of the bond forces. Using $v-u=2W$ and the
gradient expansion above, to second order in gradients,
\begin{align}
\Delta_1\equiv v_{n-1}-u_n &\simeq 2W - a v_x + \frac{a^2}{2}v_{xx},\\
\Delta_2\equiv u_{n+1}-v_n &\simeq -2W + a u_x + \frac{a^2}{2}u_{xx}.
\end{align}
The intracell bond gives $(v-u)^3=8W^3$ and $(u-v)^3=-8W^3$, while to first order in $a$
the intercell bonds give
\begin{equation}
 \begin{split}
&\Delta_1^3\simeq 8W^3-12\,aW^2v_x, \\
&\Delta_2^3\simeq -8W^3+12\,aW^2u_x .
\end{split}
\end{equation}
Inserting in Eq.~(\ref{eq:MotionNonLinear}),
\begin{equation}
 \begin{split}
m u_{tt} &= K_0(v-u) - aK_B v_x + \frac{a^2K_B}{2}v_{xx} \\
&\quad +8(\beta_A+\beta_B)W^3-12a\beta_BW^2v_x,
\end{split}
\end{equation}
and,
\begin{equation}
 \begin{split}
    m v_{tt} &= K_0(u-v) + aK_B u_x + \frac{a^2K_B}{2}u_{xx} \\
             &\quad - 8(\beta_A+\beta_B)W^3 + 12a\beta_B W^2 u_x .
 \end{split}
\end{equation}

Taking half the sum and half the difference, with $u_x=U_x-W_x$ and $v_x=U_x+W_x$,
the non-linear continuum equations become

\begin{align}
mU_{tt}
&= -aK_B W_x + \frac{a^2K_B}{2}U_{xx}
   - 12a\beta_B W^2 W_x,
\label{Eq:continuum_equation_1}
\\
mW_{tt}
&= -2K_0 W + aK_B U_x
   - \frac{a^2K_B}{2}W_{xx}
\notag\\
&\quad - 8(\beta_A+\beta_B)W^3
   + 12a\beta_B W^2 U_x.
\label{Eq:continuum_equation_2}
\end{align}

The resulting continuum equations constitute a nonlinear, dispersive, two-field hyperbolic system of Boussinesq type, describing coupled acoustic and gapped optical modes.
The terms of order $a^2$ in the cubic expansion, $6a^2W^2v_{xx}+6a^2Wv_x^2$ (and the corresponding ones in $\Delta_2^3$),
are discarded; this truncation is consistent when $W=O(1)\gg aU_x$. In the long-wavelength sector, where $W\sim aU_x$,
all terms of the cubic expansion are of the same order, see Sec.~IV. In the Appendix, we present some solitary waves solutions of the continuum model, obtained through a variational process. 
 Interestingly, the equations lead to  an undamped and unforced
cubic–quintic Duffing equation, similar to that obtained in the limit of weak coupling
without considering the continuum.

\subsection{Nature of the nonlinearities}
 
The full continuum theory contains two qualitatively distinct types of nonlinear terms.
 
\subsubsection{(A) Cubic onsite nonlinearity}
 
The first contribution is a local cubic nonlinearity of the form
\begin{equation}
-8(\beta_A + \beta_B) W^3,
\end{equation}
which corresponds to a $\phi^4$-type self-interaction field. This term arises from the local
elastic energy of each bond and represents a purely on-site nonlinear potential energy
contribution for the optical field $W(x,t)$. It cancels in the acoustic equation and appears
only in the optical one.
 
\subsubsection{(B) Derivative nonlinear terms}
 
The second contribution is a gradient-dependent nonlinear term, which appears in both equations,
\begin{equation}
-12 a \beta_B W^2 W_x=-4a\beta_B\,\partial_x\!\left(W^3\right)
\quad\text{in Eq.~(\ref{Eq:continuum_equation_1})},
\end{equation}
and,
\begin{equation}
+12 a \beta_B W^2 U_x
\quad\text{in Eq.~(\ref{Eq:continuum_equation_2})},
\end{equation}
which has a fundamentally different structure. The first one is a nonlinear flux of the optical
amplitude acting on the acoustic field, and the second one is a strain-dependent renormalization of the
optical mass; together they represent a gradient-coupled nonlinearity.
 
Such terms are characteristic of nonlinear elasticity and dispersive hydrodynamics~\cite{ElHoefer2016, PorubovAndrianov2013},
and are closely related to nonlinear Boussinesq-type systems~\cite{PorubovAndrianov2013,ChristovMauginPorubov2007}. Unlike the onsite cubic
interaction, this contribution explicitly couples amplitude and spatial variation,
leading to nonlinear transport effects.
 
\subsection{Lagrangian density}
 
The continuum equations follow from a nonrelativistic two-field Lagrangian density with derivative coupling
and quartic onsite interactions, where the acoustic and optical sectors appear as two coupled canonical fields in a dispersive nonlinear medium. Using the Pauli matrices $\sigma_j$, with $j=x,y,x$, the Lagrangian density is given by,

\begin{equation}
\begin{split}
\mathcal{L}
=&
\frac{m}{2}\,\partial_t \Phi^\mathrm{T}\partial_t \Phi
-\frac{a^2K_B}{4}\,\partial_x \Phi^\mathrm{T}\sigma_z\partial_x \Phi \\
&-\frac{aK_B}{2}\,\Phi^\mathrm{T} (i\sigma_y)\partial_x \Phi
- V(U_x,W),
\end{split}
\end{equation}
 
with
 
\begin{equation}
\Phi =
\begin{pmatrix}
U \\ W
\end{pmatrix},
\qquad
\Phi^\mathrm{T}(i\sigma_y)\partial_x\Phi=UW_x-WU_x,
\end{equation}
 
and
 
\begin{equation}
V(U_x,W)
=
K_0 W^2
+
2(\beta_A+\beta_B)W^4
-
4a\beta_B W^3 U_x .
\end{equation}
The Euler--Lagrange equations of $\mathcal L$ are exactly Eqs.~(\ref{Eq:continuum_equation_1})--(\ref{Eq:continuum_equation_2}).
Note that the gradient term has the opposite sign for $U$ and $W$ (matrix $\sigma_z$), and that
$4a\beta_BW^3W_x=a\beta_B\,\partial_x(W^4)$ would be a total derivative that gives no equation of motion:
the derivative nonlinearity comes from the coupling $W^3U_x$. The gauge-like reformulation of the continuum theory reveals a number of physically transparent structures.
 
\subsubsection{Emergent connection}
 
The mixed derivative term
\begin{equation}
\Phi^{\mathrm{T}} (i\sigma_y) \partial_x \Phi
\end{equation}
acts as an emergent connection in field space (for real fields $\Phi^{\mathrm{T}}\sigma_y\partial_x\Phi$ is purely imaginary, so the real antisymmetric matrix $i\sigma_y$ appears). In this representation, the acoustic and
optical degrees of freedom are coupled through a structure that is formally analogous to
a non-Abelian Berry connection \cite{Wilczek1984, Palumbo2021, BoschCyrill2024}, with the Pauli matrix $\sigma_y$ playing the role of a
generator in the internal two-component space.
 
\subsubsection{Acoustic--optical mixing as gauge coupling}
 
The coupling between the acoustic field $U$ and the optical field $W$ can be interpreted
as an effective gauge interaction in the continuum description. In particular, the
$U \leftrightarrow W$ mixing induced by spatial gradients behaves as a synthetic
spin--orbit coupling \cite{Zheng2024, Wang2025SpinOrbit, Gu2019SpinOrbit}, where spatial motion is locked to the internal sublattice degree
of freedom.

The nonlinear sector separates naturally into two distinct contributions. The quartic
term
\begin{equation}
W^4
\end{equation}
acts as a local $\phi^4$-type potential \cite{Campbell2019}, governing onsite nonlinear elasticity of the
optical mode \cite{PorubovAeroMaugin2009}. In contrast, the derivative nonlinear term
\begin{equation}
W^3 U_x
\end{equation}
couples the optical amplitude with the acoustic strain. It generates the nonlinear current
$-4a\beta_B\partial_x(W^3)$ in the acoustic equation and the strain-dependent term
$12a\beta_BW^2U_x$ in the optical one, leading to nonlinear transport effects that are absent in purely local field
theories.

\section{Long-wavelength limit}
 
Eliminating the fast optical field yields an effective nonlinear wave equation
\begin{equation}
mU_{tt}=mc^2U_{xx}+\gamma U_x^2 U_{xx}+\lambda U_{xxxx}+\cdots.
\end{equation}
Indeed, inserting $W\simeq aK_BU_x/(2K_0)$ in Eq.~(\ref{Eq:continuum_linear_1}) gives
$mU_{tt}=\frac{a^2K_B}{2}\left(1-\frac{K_B}{K_0}\right)U_{xx}=mc^2U_{xx}$.
The dispersive coefficient follows from the $q^4$ term of the exact dispersion relation,
\begin{equation}
\lambda=\frac{a^4K_AK_B\left(K_A^2+K_B^2-K_AK_B\right)}{24\,(K_A+K_B)^3}.
\end{equation}
 
In this regime $W\sim aU_x$, so all terms in the cubic expansion of the bond stretches are of the same
order $a^3U_x^3$ and the first-order truncation of Eqs.~(\ref{Eq:continuum_equation_1})--(\ref{Eq:continuum_equation_2}) is not sufficient.
The nonlinear coefficient is obtained from the force balance of the two bonds in series. With
intracell stretch $d_A=2W$ and intercell stretch $d_B=aU_x-2W$,
\begin{equation}
F_A(d_A)=F_B(d_B)\equiv F,\qquad d_A+d_B=aU_x ,
\end{equation}
which gives, to cubic order,
\begin{equation}
F=\frac{K_AK_B}{K_0}\,aU_x+\left(\frac{K_AK_B}{K_0}\right)^4\left(\frac{\beta_A}{K_A^4}+\frac{\beta_B}{K_B^4}\right)a^3U_x^3 ,
\end{equation}
and the acoustic equation is $mU_{tt}=\frac{a}{2}\partial_xF$. Hence
\begin{equation}
\gamma=\frac{3a^4\left(\beta_AK_B^4+\beta_BK_A^4\right)}{2\,(K_A+K_B)^4}.
\end{equation}
 
\section{Continuum Dirac limit}
 
The most important regime arises near the Brillouin-zone edge $q=\pi/a$. We factor out the fast spatial oscillation $(-1)^n$ and the carrier frequency $\omega_0$:
\begin{equation}
u_n = (-1)^n \psi_A, \qquad v_n = (-1)^n \psi_B,
\end{equation}
where,
\begin{equation}
    \psi_{A,B}(x,t)=\Phi_{A,B}(x,t)\,e^{-i\omega_0t}+\mathrm{c.c.},
\end{equation}
with slowly varying envelopes $\Phi_{A,B}$.
Defining
\begin{equation}
K=\frac{K_A+K_B}{2}, \qquad \delta=\frac{K_A-K_B}{2},
\end{equation}
and removing the optical frequency $\omega_0^2=2K/m$ (neglecting $\partial_t^2\Phi$ and second derivatives), we obtain
for the pure linear theory,
\begin{equation}
i\partial_t
\begin{pmatrix}\Phi_A\\\Phi_B\end{pmatrix}
=
\left[-iv\sigma_y\partial_x + M\sigma_x\right]
\begin{pmatrix}\Phi_A\\\Phi_B\end{pmatrix},
\end{equation}
with
\begin{equation}
v=\frac{aK}{2m\omega_0}, \qquad M=-\frac{\delta}{m\omega_0}=\frac{K_B-K_A}{2m\omega_0}.
\end{equation}
The spectrum is $\omega=\omega_0\pm\sqrt{v^2k^2+M^2}$, with $k=q-\pi/a$, in agreement with the expansion of the exact branches~(9) near the zone edge.

This is the continuum SSH Dirac Hamiltonian. It is a Jackiw--Rebbi Hamiltonian that supports topological solitons (zero modes at the domain walls where $M$ changes sign) \cite{jackiw1976, RajabpoorAlisepahi2023, Cooper2010}, which have also been found in moiré graphene systems and identified with flat-band electronic modes associated with superconductivity and charge fractionalization \cite{Elias_2023,Andrade_2025,Elias2026}. Notice that the present result is also in agreement with the observation made in a previous work that the flat bands of the electronic problem in twisted bilayer graphene at magic angles correspond to topological flexible mechanical modes, known as floppy modes in the context of rigidity theory \cite{Naumis_2011}, once a supersymmetry transformation is applied \cite{Naumis2021}.

Notice that a rigidity transition occurs here when the fields are decoupled by setting $K_B=0$. In this limit, the system consists of $N/2$ dimers, and the field $U$ represents the center-of-mass coordinate of each dimer. Consequently, the associated frequencies are zero, resulting in a floppy-mode peak with a degeneracy of $N/2$. The optical field, on the other hand, corresponds to the stretching mode of each dimer. Therefore, as the system evolves from $K_A=K_B$ to $K_B=0$, it undergoes a transition from an isostatic to a flexible mechanical system \cite{Naumis_2011, toledo2018escape}.

\section{Nonlinear Dirac theory}
 
Near the zone edge the bond stretches are not $v-u$ for both bonds. Using the factorization above,
\begin{equation}
\begin{split}
&v_n-u_n=(-1)^n(\psi_B-\psi_A), \\
&v_{n-1}-u_n\simeq u_{n+1}-v_n\simeq-(-1)^n(\psi_A+\psi_B),
\end{split}
\end{equation}
so the intracell nonlinearity $\beta_A$ acts on $\psi_B-\psi_A$ and the intercell one $\beta_B$ on $\psi_A+\psi_B$.
In the rotating-wave approximation $(Xe^{-i\omega_0t}+\mathrm{c.c.})^3\to3|X|^2Xe^{-i\omega_0t}$.
Defining $\Phi_\pm=\Phi_A\pm\Phi_B$, and
\begin{equation}
g_{A,B}=\frac{3\beta_{A,B}}{2m\omega_0},
\end{equation}
we obtain
\begin{align}
i\partial_t \Phi_A &= -v\,\partial_x \Phi_B + M\Phi_B
+ g_B|\Phi_+|^2\Phi_+ + g_A|\Phi_-|^2\Phi_-, \\
i\partial_t \Phi_B &= +v\,\partial_x \Phi_A + M\Phi_A
+ g_B|\Phi_+|^2\Phi_+ - g_A|\Phi_-|^2\Phi_- ,
\end{align}
where the kinetic part is the explicit form of $-iv\sigma_y\partial_x+M\sigma_x$.
In spinor form,
\begin{equation}
i\partial_t\Psi=
\left(-iv\sigma_y\partial_x + M\sigma_x\right)\Psi
+ \mathcal{N}(\Psi),
\quad \Psi=\begin{pmatrix}\Phi_A\\\Phi_B\end{pmatrix}.\\
\end{equation}
where $\mathcal{N}(\Psi)$ is the nonlinear part.

\section{Soler limit}
 
In the eigenbasis of $\sigma_x$, $\chi=(\Phi_A+\Phi_B,\;\Phi_B-\Phi_A)^{\mathrm T}/\sqrt2$, the mass term becomes diagonal and
\begin{equation}
\begin{split}
&i\partial_t\chi=\left(-iv\sigma_y\partial_x + M\sigma_z\right)\chi
+(G_B|\chi_1|^2\chi_1, G_A|\chi_2|^2\chi_2)^{\mathrm T}, \\
&G_{A,B}=\frac{6\beta_{A,B}}{m\omega_0}.
\end{split}
\end{equation}
For $\beta_A=\beta_B=\beta$ ($G_A=G_B=G$) the nonlinearity splits as
\begin{equation}
\mathcal N(\chi)=\frac{G}{2}\left(\chi^\dagger\chi\right)\chi+\frac{G}{2}\left(\bar\chi\chi\right)\sigma_z\chi,
\qquad \bar\chi=\chi^\dagger\sigma_z .
\end{equation}
The second term is the Lorentz-scalar (Soler) interaction ~\cite{soler1970, CuevasMaraver2018Solitary}, while the first one is a density--density
(Thirring-type) term \cite{Thirring1958,Coleman1975}, which is not Lorentz scalar. We observe that the previous equation is a generalized Dirac non linear equation. In fact, it only the scalar part is considered, and using
$\gamma^0=\sigma_z$, $\gamma^1=-i\sigma_x$ (so that $\gamma^0\gamma^1=\sigma_y$) and units with $v=1$, we obtain the usual nonlinear Dirac (Soler) equation 
\begin{equation}
\begin{split}    
i\gamma^\mu \partial_\mu& \Psi - M\Psi + g(\bar\Psi\Psi)\Psi=0,\\
&g=-\frac{G}{2}=-\frac{3\beta}{m\omega_0}.
\end{split}
\end{equation}
The fact that the Thirring term is not Lorentz invariant does not, by itself, justify dropping it from the original physical equations. Lorentz invariance is a property imposed on the reduced effective model, not necessarily a symmetry of the underlying lattice system. Therefore, the obtained equation that contains the Soler and Thirring terms is more general than the Soler equation,
 
\section{Conclusion}
 
We have shown that a nonlinear dimerized mechanical SSH chain generates, in the continuum limit, a nonlinear Dirac field theory whose cubic interaction contains a Soler-type scalar term (together with a density--density term). This provides a mechanical realization of interacting relativistic-like solitons, including nonlinear Jackiw--Rebbi states and self-trapped topological excitations.

\section*{Data Availability}

The data supporting the findings of this study are available
within the article. The dispersion relations and solitary-wave
profiles can be reproduced from the equations and parameter
values provided in the main text and the Appendix.

\appendix

\section{Solitary waves of the continuum model}
\label{app:solitary-wave}
 
This appendix describes a variational route to solitary profiles and
then constructs a traveling solution of the original continuum model \cite{Hasegawa2003OpticalSolitons}.
The distinction between an exact profile of the continuum equations
and a prediction of a trial-function approximation is important here.
The system \ref{Eq:continuum_equation_1}-\ref{Eq:continuum_equation_2} is a system of the form:
\begin{align}
 U_{tt}&-a_1U_{xx}+a_2W_x+a_3W^2W_x=0,\label{eq:app_U}\\
 W_{tt}&+a_1W_{xx}+b_1W+b_2W^3-a_2U_x-a_3W^2U_x=0,\label{eq:app_W}
\end{align}
with
\begin{equation}
\begin{split}
 a_1=\frac{a^2K_B}{2m},\quad a_2=\frac{aK_B}{m},\quad a_3=\frac{12a\beta_B}{m},\\
 b_1=\frac{2K_0}{m},\quad b_2=\frac{8(\beta_A+\beta_B)}{m}.
\end{split}
\end{equation}
 Here $a$ is the unit-cell period. Both the sign of $W_{xx}$ and the reciprocal nonlinear coupling in Eq.~\eqref{eq:app_W} are essential. We assume $a_1>0$ and $b_1>0$.

A normalized Lagrangian density is
\begin{equation}
\begin{split}
 \mathcal L=\frac{U_t^2+W_t^2}{2}
 -\frac{a_1}{2}(U_x^2-W_x^2)-\frac{b_1}{2}W^2-\frac{b_2}{4}W^4\\
 -a_2UW_x+\frac{a_3}{3}W^3U_x.
\label{eq:app_L}
\end{split}
\end{equation}
Its Euler--Lagrange equations are exactly Eqs.~\eqref{eq:app_U}--\eqref{eq:app_W}, no nonvariational residual needs to be appended. The physical action density inherited from the lattice has an overall factor $2m/a$, which does not affect these equations.

For trial fields parametrized by collective coordinates $Q_j(t)$, an averaged Lagrangian $L=\int_{\mathbb R}\mathcal L\,dx$ gives $d(\partial L/\partial\dot Q_j)/dt-\partial L/\partial Q_j=0$, provided the action is finite and boundary terms vanish. Two Gaussians even about the same center make both coupling integrals $\int UW_x\,dx$ and $\int W^3U_x\,dx$ vanish. Such an ansatz cannot resolve their effect on amplitude and width. An even optical pulse accompanied by an acoustic displacement step has the appropriate parity. We obtain this profile directly, without a trial-function approximation.

\subsection{Traveling-wave reduction}
Let $U(x,t)=u(\xi)$ and $W(x,t)=w(\xi)$, where $\xi=x-Vt$. We use $V$ for the pulse velocity to distinguish it from other velocities in the text. Impose $w,w',u'\to0$ as $|\xi|\to\infty$, and define
\begin{equation}
 D=V^2-a_1\ne0,\qquad S=V^2+a_1>0.
\end{equation}
Integrating Eq.~\eqref{eq:app_U} once fixes the integration constant to zero:
\begin{equation}
 u'=-\frac{a_2w+(a_3/3)w^3}{D}.\label{eq:app_strain}
\end{equation}
Substitution into Eq.~\eqref{eq:app_W} gives a cubic--quintic profile equation,
\begin{equation}
\begin{split} 
 &Sw''+Cw+Ew^3+Gw^5=0,\\
 &C=b_1+\frac{a_2^2}{D},\quad E=b_2+\frac{4a_2a_3}{3D},\quad G=\frac{a_3^2}{3D}.
 \end{split}
 \label{eq:app_profile}
\end{equation}
This is a Duffing-type equation in the traveling coordinate $\xi$ \cite{Belendez2016CubicQuintic}. Multiplication by $w'$ yields
\begin{equation}
 \frac S2(w')^2+\frac C2w^2+\frac E4w^4+\frac G6w^6=0.\label{eq:app_energy}
\end{equation}
The zero on the right follows from the localized boundary conditions.

\subsection{Localized profile and displacement step}
A sufficient parameter regime for a smooth bright pulse is
\begin{equation}
 D<0,\qquad C<0,\qquad E>0,\qquad E^2>\frac{16}{3}CG.\label{eq:app_exist}
\end{equation}
In this regime $G\le0$. Setting $z=w^{-2}$ in Eq.~\eqref{eq:app_energy} gives
\begin{equation}
 (z')^2=4\kappa^2z^2-\frac{2E}{S}z-\frac{4G}{3S},\qquad \kappa=\sqrt{-C/S}.
\end{equation}
The positive pulse centered at $\xi=0$ is
\begin{equation}
\begin{split}
&w(\xi)=\frac{1}{\sqrt{P+Q\cosh(2\kappa\xi)}},\\
 &P=-\frac{E}{4C},\quad Q=\sqrt{P^2-\frac{G}{3C}}.\label{eq:app_exact}
\end{split}
\end{equation}
Conditions~\eqref{eq:app_exist} ensure $P>0$ and $Q>0$, so the denominator never vanishes. Its amplitude is $A=(P+Q)^{-1/2}$; its far-field decay length is $\kappa^{-1}$. As $G\to0$ with $C<0$ and $E>0$ fixed, this expression tends to $w=A\,\mathrm{sech}(\kappa\xi)$, with $A^2=-2C/E$.

The acoustic field follows by quadrature:
\begin{align}
 u(\xi)&=u_- -\frac1D\int_{-\infty}^{\xi}\left[a_2w(s)+\frac{a_3}{3}w^3(s)\right]ds,\label{eq:app_u}\\
 \Delta u\equiv u_+-u_-&=-\frac1D\int_{-\infty}^{\infty}\left[a_2w(s)+\frac{a_3}{3}w^3(s)\right]ds.\label{eq:app_jump}
\end{align}
Thus $w$ and $u'$ are localized, whereas $u$ connects two constant displacement levels. The strain is a linear-plus-cubic function of $w$, not exactly proportional to it. For $a_2,a_3>0$ and $D<0$, the positive pulse produces a positive strain and an increasing displacement step. The transformation $(u-u_-,w)\mapsto(-(u-u_-),-w)$ gives the opposite-polarity solution.

\subsection{An explicit example}
\begin{figure}[htbp]
 \centering
 \includegraphics[width=\linewidth]{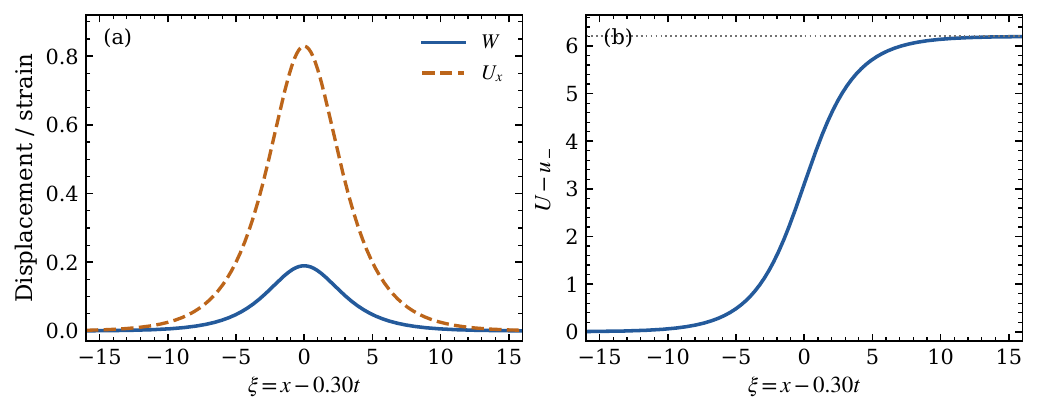}
 \caption{Exact traveling solution of the truncated continuum equations for the parameters in Eq.~\eqref{eq:app_example}. (a) Localized optical displacement $W$ (blue, solid) and acoustic strain $U_x$ (orange, dashed). (b) Acoustic displacement relative to its left asymptote, $U-u_-$, obtained by quadrature; the dotted line marks $\Delta u=6.201589331$. Both panels use $\xi=x-0.30t$. The curves illustrate a solution of the truncated field model and do not constitute a discrete-lattice simulation.}
 \label{fig:app_pulse}
\end{figure}

In dimensionless variables, take
\begin{equation}
\begin{split}
 &a_1=0.15,\qquad a_2=0.25,\qquad a_3=1,\\
 &b_1=1,\qquad b_2=8,\qquad V=0.30.
 \end{split}
 \label{eq:app_example}
\end{equation}
These values give
\begin{equation}
\begin{split}
 D=-\frac3{50},\quad S=\frac6{25},\quad C=-\frac1{24},\quad E=\frac{22}{9},\\
 G=-\frac{50}{9},\quad \kappa=\frac5{12},\quad P=\frac{44}{3},\quad Q=\frac{16\sqrt6}{3}.
 \end{split}
\end{equation}
All inequalities in Eq.~\eqref{eq:app_exist} hold. In particular, $E^2-16CG/3=128/27>0$. The solution reads
\begin{equation}
\begin{split}
 &W(x,t)=\left[\frac{44}{3}+\frac{16\sqrt6}{3}\cosh\!\left(\frac56(x-0.30t)\right)\right]^{-1/2},\\
 &U_x=\frac{25}{6}W+\frac{50}{9}W^3.\label{eq:app_example_fields}
\end{split}
\end{equation}
With $u_-=0$, numerical quadrature of Eq.~\eqref{eq:app_u} yields
\begin{equation}
\begin{split}
 &W(0,0)=0.1898979486,\quad U_x(0,0)=0.8292856399,\\ &\Delta u=6.201589331.
\end{split}
\end{equation}
Figure~\ref{fig:app_pulse} uses precisely these parameters and equations. Substitution into Eqs.~\eqref{eq:app_U}--\eqref{eq:app_W} gives identically zero residual. Shape-preserving translation follows from the traveling ansatz.

\bibliographystyle{apsrev4-2}
\bibliography{BIB_SSH_Continuum}

\end{document}